\documentclass[superscriptaddress,twocolumn,showpacs,amsmath,amssymb,prl]{revtex4}

\usepackage{graphicx}% Include figure files
\usepackage{dcolumn}% Align table columns on decimal point
\usepackage{bm}% bold math

\begin{document}

%\title{Approaching ground state Born-Oppenheimer molecular dynamics\\
%  in a modified time-dependent density functional theory approach}

%\title{New Ehrenfest TDDFT approach for efficient large-scale ab initio BO Molecular Dynamics}
%\title{An efficient time-dependent density functional formalism for
%  large scale ab-initio molecular dynamics}
%\title{Modified time-dependent density functional formalism for
%  efficient large scale ab-initio molecular dynamics}

\title{Efficient formalism for large scale ab initio molecular dynamics\\  
  based on time-dependent density functional theory}

\author{J. L. Alonso}
\affiliation{Departamento de F{\'{\i}}sica Te\'orica, Universidad de Zaragoza, Pedro Cerbuna 12, E-50009 Zaragoza, Spain.}
\affiliation{Instituto de Biocomputaci\'on y F{\'{\i}}sica de Sistemas Complejos (BIFI).}

\author{X. Andrade}
\affiliation{European Theoretical Spectroscopy Facility, Departamento 
  de F{\'{\i}}sica de Materiales, Universidad del Pa{\'{i}}s Vasco,
  Centro Mixto CSIC-UPV, and DIPC, Edificio Korta,
  Av. Tolosa 72, E-20018 San Sebasti\'an,
  Spain}

\author{P. Echenique}
\affiliation{Departamento de F{\'{\i}}sica Te\'orica, Universidad de Zaragoza, Pedro Cerbuna 12, E-50009 Zaragoza, Spain.}
\affiliation{Instituto de Biocomputaci\'on y F{\'{\i}}sica de Sistemas Complejos (BIFI).}

\author{F. Falceto}
\affiliation{Departamento de F{\'{\i}}sica Te\'orica, Universidad de Zaragoza, Pedro Cerbuna 12, E-50009 Zaragoza, Spain.}
\affiliation{Instituto de Biocomputaci\'on y F{\'{\i}}sica de Sistemas Complejos (BIFI).}

\author{D. Prada-Gracia}
\affiliation{Departamento de F{\'{\i}}sica de la Materia
  Condensada, Universidad de Zaragoza.}
\affiliation{Instituto de Biocomputaci\'on y F{\'{\i}}sica de Sistemas Complejos (BIFI).}

\author{A. Rubio}
\affiliation{European Theoretical Spectroscopy Facility, Departamento 
  de F{\'{\i}}sica de Materiales, Universidad del Pa{\'{i}}s Vasco,
  Centro Mixto CSIC-UPV, and DIPC, Edificio Korta,
  Av. Tolosa 72, E-20018 San Sebasti\'an,
  Spain}

\date{\today}

\begin{abstract}

  A new ``on the fly'' method to perform Born-Oppenheimer ab initio
  molecular dynamics (AIMD) is presented. Inspired by Ehrenfest dynamics
  in time-dependent density functional theory, the electronic orbitals
  are evolved by a Schr{\"o}dinger-like equation, where the orbital
  time derivative is multiplied by a parameter. This parameter
  controls the time scale of the fictitious electronic motion and
  speeds up the calculations with respect to standard Ehrenfest
  dynamics. In contrast to other methods, wave function orthogonality
  needs not be imposed as it is automatically preserved, which is of
  paramount relevance for large scale AIMD simulations.

\end{abstract}


\maketitle

Ab initio molecular dynamics (AIMD) on the ground state
Born-Oppenheimer (gsBOMD) potential energy surface for the nuclei has
become a standard tool for simulating the conformational behaviour of
molecules, bio- and nano-structures and condensed matter systems from
first principles~\cite{Mar2000TR}. However, gsBOMD (in the
DFT~\cite{Koh1965PR} picture) requires that the Kohn-Sham (KS) energy
functional be minimized for each value of the nuclei positions. As
this minimization can be very demanding, Car and Parrinello
(CP)~\cite{Car1985PRL} proposed an elegant and efficient
``on~the~fly'' scheme in which the KS orbitals are propagated with a
fictitious dynamics that mimics gsBOMD. The CP method has had a
tremendous impact in many scientific
areas~\cite{And2005CPC,Tuc2002JPCM}. Nevertheless, the numerical cost
of AIMD hinders the application of the method to large scale
simulations, such as those of interest in biochemistry or material
science. Recently, new methods that allow larger systems and longer
simulation times to be studied have been reported \cite{Kuhne2007},
but the cost associated with the wave function orthogonalization
is still a potential bottleneck for both gsBOMD and CP.

Time-dependent density functional theory (TDDFT) \cite{Mar2006BOOK,
  Run1984PRL} has been for a long time recognized as an
orthogonalization-free alternative for both ground state
\cite{Selloni} and excited state AIMD. In its simplest implementation,
Ehrenfest TDDFT, the ions are treated classically following
electronic Hellmann-Feynman forces. For systems where the gap between
the ground and the first excited state is large, Ehrenfest tends to
gsBOMD and can mimic adiabatic dynamics~\cite{Mar2000TR}. However, the
rapid movement of the electrons in TDDFT requires the use of a very
small time step, which, in many occasions, renders its numerical
application non-practical~\cite{The1992PRB}.

In this letter, we borrow some of the ideas of CP and introduce a new
TDDFT Ehrenfest dynamics that reduces the cost of AIMD simulations
while keeping the accuracy of the results in tolerable levels, similar
to CP. The whole scheme can be obtained from the following Lagrangian
(atomic units are used throughout this paper):
\begin{equation}
\label{LZ}
\mathcal{L}= i\,\frac{\mu}{2} \sum_{j=1}^{N} \int \left( \phi_j^*\dot{\phi}_j
 - \dot{\phi}_j^*\phi_j \right)
\mathrm{d}{\bm r} + K_I - E[\phi,R] \;,
\end{equation}
where $K_I=\frac{1}{2}\sum_{I}M_I\dot{{\bm R}}_I\!\cdot\!\dot{{\bm
    R}}_I$ is the kinetic energy of the nuclei, $M_I$ their masses and
\(E\) the KS energy. Note that the major modification with respect to
TDDFT is the scaling of the electronic velocities by a parameter $\mu$
(TDDFT is recovered when \(\mu=1\)). We show in what
follows that, in the \(\mu\to0\) limit, the trajectories of the
system approach gsBOMD, and practical calculations can be done
for values of \(\mu\gg1\), thus allowing for more efficient
implementations than TDDFT while retaining its advantageous
properties: the conservation of the total energy and of the
orthogonality of the orbitals. Also, from the computational point of
view, the new scheme is simple and can be easily incorporated into
existing codes.

The equations of motion obtained from~(\ref{LZ}) for the
electronic~($\phi_j$) and nuclear ($\bm{R}_I$) degrees of freedom are:
\begin{subequations}
\label{EoMZ}
\begin{align}
  i \,\mu \dot{\phi}_j &=\frac{\delta E[\phi,R]}{\delta \phi_j^*} =
  -\frac{1}{2}\nabla^2 \phi_j + v_{\mathrm{eff}}({\bm r, t})
  \phi_j \;, \label{EoMZa}\\
  M_I \ddot{{\bm R}}_I &= - {\boldsymbol \nabla}_I E[\phi,R] \ , \label{EoMZc}
\end{align}
\end{subequations}
where \(v_{\mathrm{eff}}\) is the time-dependent KS effective
potential.

In contrast to CP, the new dynamics conserves the physical energy
$E_{\mathrm{phys}}:=K_I+E[\phi,R]$ as well as the scalar product among
the orbitals $\phi_j$. The first is a direct consequence of
$\mathcal{L}$ being linear in the velocities $\dot{\phi}_j$ and
$\dot{\phi}^*_j$, and not depending explicitly on $t$. The
conservation of the scalar product requires more attention due to the
nonlinear character of the term $\delta E/\delta \phi_j^*$.  To prove
it, note that $E[\phi,R]$ is invariant under any unitary
transformation mixing the orbitals $\phi \rightarrow U\phi$, with
$U=e^{-i(\varepsilon/\mu)A}$, being $A$ an $N \times N$ Hermitian
matrix. From this invariance and eq.~(\ref{EoMZ}), we have
\begin{eqnarray}
\label{EscProd}
\lefteqn{\frac{\mathrm{d}}{\mathrm{d}t}\int
A_{jk}\phi_j^*\phi_k\,\mathrm{d}{\bm r}=
\int \left(
A_{jk}\dot{\phi}_j^*\phi_k +
A_{jk}\phi_j^*\dot{\phi}_k\right)\mathrm{d}{\bm r}}\nonumber\\
&& = -\frac{i}{\mu}\int\left(
A_{jk}\frac{\delta E}{\delta \phi_j}\phi_k -
A_{jk}\phi_j^*\frac{\delta E}{\delta \phi^*_k}
\right)\mathrm{d}{\bm r}\nonumber \\ 
&& = \frac{\mathrm{d}}{\mathrm{d}\varepsilon}
E\left[e^{-i{\scriptstyle \frac{\varepsilon}{\mu}}A}\phi, R \right]
\bigg|_{\varepsilon = 0} = 0
\;.
\end{eqnarray}
Now, since $\int A_{jk} \phi_j^*\phi_k\,\mathrm{d}{\bm r}$ is constant
for all $A=A^\dagger$, the scalar product of any pair $\phi_j$,
$\phi_k$ is a constant as well. Hence, if we start from an orthonormal
set, we will not have to reorthonormalize the orbitals during the MD
simulation. Numerically, this means that the formal scaling of
the new scheme is quadratic with the number of atoms, while for CP and
gsBOMD it is cubic~\footnote{Assuming no linear
  scaling techniques are used. Note that these techniques could also
  be combined with our method, where the orthogonalization-free
  propagation would be reflected in a smaller scaling prefactor.} due
to the orthogonalization. In addition, the time propagation is
naturally parallelizable by distributing the orbitals among different
processors, as the evolution of each orbital is almost independent
from the others'.

An important question is whether the new method reproduces gsBOMD. We
show that the \(\mu\to0\) limit accounts for this solution. To do so,
we recall that the BO Lagrangian reads
\begin{equation}
\mathcal{L}_{\mathrm{BO}} = K_I\! -\! E[\phi,R]
+\! \sum_{jk} \Lambda^{\mathrm{BO}}_{jk} \left( \int\! \phi_j^*\phi_k
\,\mathrm{d}{\bm r}\! -\!\delta_{jk} \right)\,, \label{LBO}
\end{equation}
where \(\Lambda^{\mathrm{BO}}_{jk}\) are the Lagrange multipliers
which ensure the orthonormality of the orbitals. Clearly, as the
orthonormality is automatically satisfied by the propagator in our
approach, the limit \(\mu\to0\) gives the BO Lagrangian without the
\(\Lambda^{\mathrm{BO}}\) term. Note, however, that one could have
started from a different Lagrangian \(\mathcal{L}^\prime =\mathcal{L}
+ \sum_{jk}\Lambda_{jk} \big(\int \phi^*_j\phi_k\,\mathrm{d}{\bm r} -
\delta_{jk}\big)\) for which the \(\mu\to0\) limit is
\(\mathcal{L}_{\mathrm{BO}}\), and then, using (for \(\mu\neq 0\)) the
gauge symmetry of \(\mathcal{L}^\prime\)
(\(\phi^\prime=\mathrm{e}^{iA}\phi\) and \(\Lambda^\prime =
\mathrm{e}^{iA} \Lambda \mathrm{e}^{-iA} - i \mu \mathrm{e}^{iA}
{\textstyle \frac{\mathrm{d}}{\mathrm{d}t}} \mathrm{e}^{-iA}\)), where
\(A\) is a time-dependent Hermitian matrix, one can send
\(\Lambda^\prime\) to zero recovering the dynamics of
\(\mathcal{L}\). This limit has a simple physical interpretation. The
effect of \(\mu\) is to scale the TDDFT excitation energies by a
\(1/\mu\) factor. So for \(\mu > 1\) the gap of the artificial system
is decreased, increasing the non-adiabatic coupling, while for small
values \(\mu\) the excited states are pushed up in energy forcing the
system to stay in the adiabatic regime \footnote{This sets the maximum
  value of \(\mu\) as the electronic gap divided by the highest
  vibration energy in the system}.

Next, to provide an estimation of the performance improvements of our
method in comparison with Ehrenfest dynamics, we write the left hand
side of~(\ref{EoMZa}) as \(\mu (d\phi/dt) = d\phi/dt_e\).  With this
transformation, (\ref{EoMZa}) can be seen as a standard TDDFT
propagation, and we find that the maximum time step for our method in
terms of \(\mu\) is $\Delta{}t=\mu\Delta{}t_e$, where \(\Delta{}t_e\)
is the maximum electronic time step, determined by the system and the
propagation scheme. In the case of CP, on the other hand,
\(\Delta{}t\propto\sqrt{\mu_\mathrm{CP}}\).  Note, however, that this
difference does not imply anything about the relative performance of
both methods, since the two parameters are not directly comparable
(e.g., they have different dimensions). The dependence of the accuracy
on $\mu$ must also be taken into account, as we show
later. Additionally, the ionic motion imposes a constraint in the
maximum value of \(\Delta{}t\), but usually this limit is much higher.

Now, although our method approaches the reference gsBOMD as
\(\mu\to0\), this limit is not practical from a numerical point of
view because it implies a time step $\Delta t \to 0$. But, as
\(\mu=1\) is already close to gsBOMD for large gap systems, we shall
mainly focus on how close we can stay to this limit for
\(\mu\gg1\). In this regime, numerical simulations are in principle
$\mu$ times faster than standard TDDFT, so we first made a detailed
study of how large can \(\mu\) be in CP and Ehrenfest for a simple
2-band model (see suplementary material~\cite{supp}), to conclude that
the new scheme shows a performance similar to CP.

To further investigate in real systems the efficiency of this new
approach, we implemented it, together with CP, in the first principle
Octopus code~\cite{Cas2006PSSB}. For Ehrenfest dynamics, the
Approximated Enforced Time Reversal Symmetry method~\cite{Cas2004JCP}
is used to propagate the electronic wave functions. In the case of CP
the electronic part is integrated by a RATTLE/Velocity Verlet
algorithm described in Ref.~\onlinecite{Tuckerman1994}. In both cases,
velocity Verlet algorithm is used for the ionic equations of
motion. The ions are represented using norm-conserving
pseudo-potentials and the exchange correlation term is approximated by
the Adiabatic LDA functional.

With respect to implementation there are some differences to
remark. For Ehrenfest, the propagation must be performed using complex
wave functions while for CP it can be performed using real wave
functions for finite systems or for gamma point super-cell
calculations in periodic systems. Also, due to the second order dynamic
of CP, two sets of wave functions must be propagated while only one is
needed for Ehrenfest. Finally, in the velocity Verlet algorithm, a
temporary third set of wave functions is required to store the
previous time step.

In parallel architectures, CP methods are known to scale very well
based on domain descomposition~\cite{Parallel-CPMD}. This also applies
to Ehrenfest dynamics and, on top of that, we can add a new level of
parallelization by distributing groups of different states among
processors. As the evolution of each state is independent, this is a
very effective approach where communication is only required to
calculate quantities that involve sums over all states, like the
density or the forces. As these operations are performed only once per
time step, it can scale efficiently even over slow
interconnections. In the case of CP, due to orthogonalization between
states, this parallelization scheme is more complex to
implement~\cite{Lorenzen} and requires much more communication.

The first real system we simulated was the Nitrogen molecule (see
supplementary material~\cite{supp}). We observed that, for \(\mu=20\), the
simulation remains steadily close to the BO potential energy surface,
and there is only a \(3.4\%\) deviation of the vibrational
frequency. For \(\mu=30\) the system starts to strongly separate from
the gsBO surface by mixing with higher BO surfaces. 

\begin{figure}
\centering
\includegraphics*[width=7.8cm]{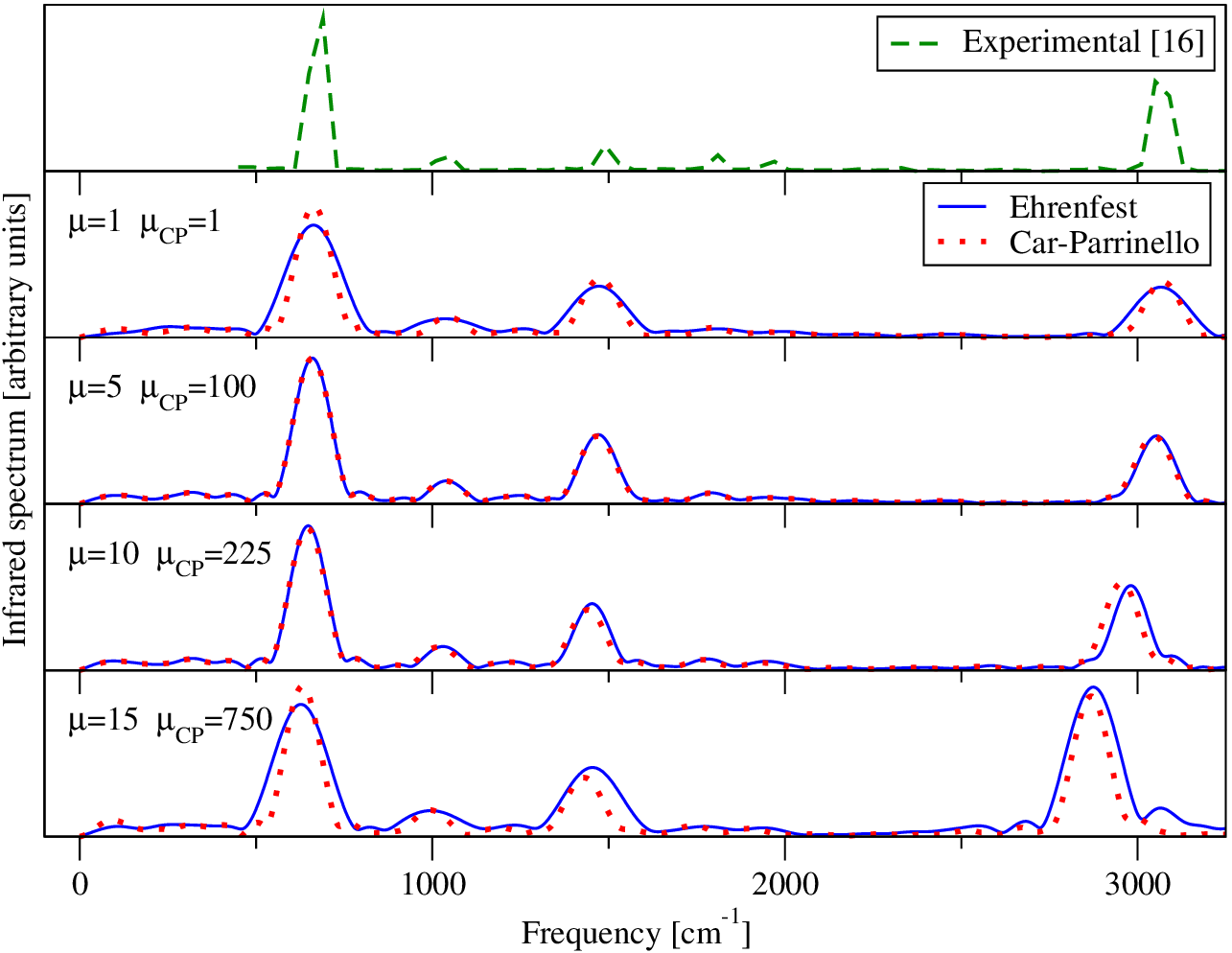}
\caption{\label{benzene} Calculated infrared spectrum for benzene for
  different values of \(\mu\), compared to CP dynamics and to
  experiment~\cite{benzene_experimental}.}
\end{figure}

Next, we applied the method to the benzene molecule. We set-up the
atoms in the equilibrium geometry with a random Maxwell-Boltzmann
distribution for \(300^\circ\mbox{K}\). Each run was propagated for a
period of time of \(\sim400\) [fs] with a time step of
\(\mu\times0.001\) [fs] (that provides a reasonable convergence in the
spectra). Vibrational frequencies were obtained from the Fourier
transform of the velocity auto-correlation function. In table
\ref{benz_freq}, we show some low, medium and high frequencies of
benzene as a function of \(\mu\). The general trend is a red-shift of
the frequencies with a maximum deviation of 7\(\%\) for
\(\mu=15\). Still, to make a direct comparison with experiment, we
computed the infrared spectra as the Fourier transform of the
electronic dipole operator. In Fig.~\ref{benzene}, we show how the
spectra changes with \(\mu\). For large \(\mu\), besides the
red-shift, spurious peaks appear above the higher vibrational
frequency (not shown). We performed equivalent CP calculations for
different values of \(\mu_\mathrm{CP}\), and found that, as shown in
Fig.~\ref{benzene}, it is possible to compare the physical error
induced in both methods and establish and a relation between \(\mu\)
and \(\mu_\mathrm{CP}\).

\begin{table}
\begin{tabular}{lrrrrrr}

\hline
\(\mu=1\)    & 398 & 961 & 1209 & 1623 & 3058 \\
\(\mu=5\)    & 396 & 958 & 1204 & 1620 & 3040 \\
\(\mu=10\)   & 391 & 928 & 1185 & 1611 & 2969 \\
\(\mu=15\)   & 381 & 938 & 1181 & 1597 & 2862 \\
\hline
\end{tabular}

\caption{\label{benz_freq} 
  Selected vibrational frequencies (in cm\(^{-1}\)) for the benzene molecule,
  obtained using different values of \(\mu\).
}
\end{table}

\begin{figure}[ht!]
\centering
\includegraphics*[scale=0.39]{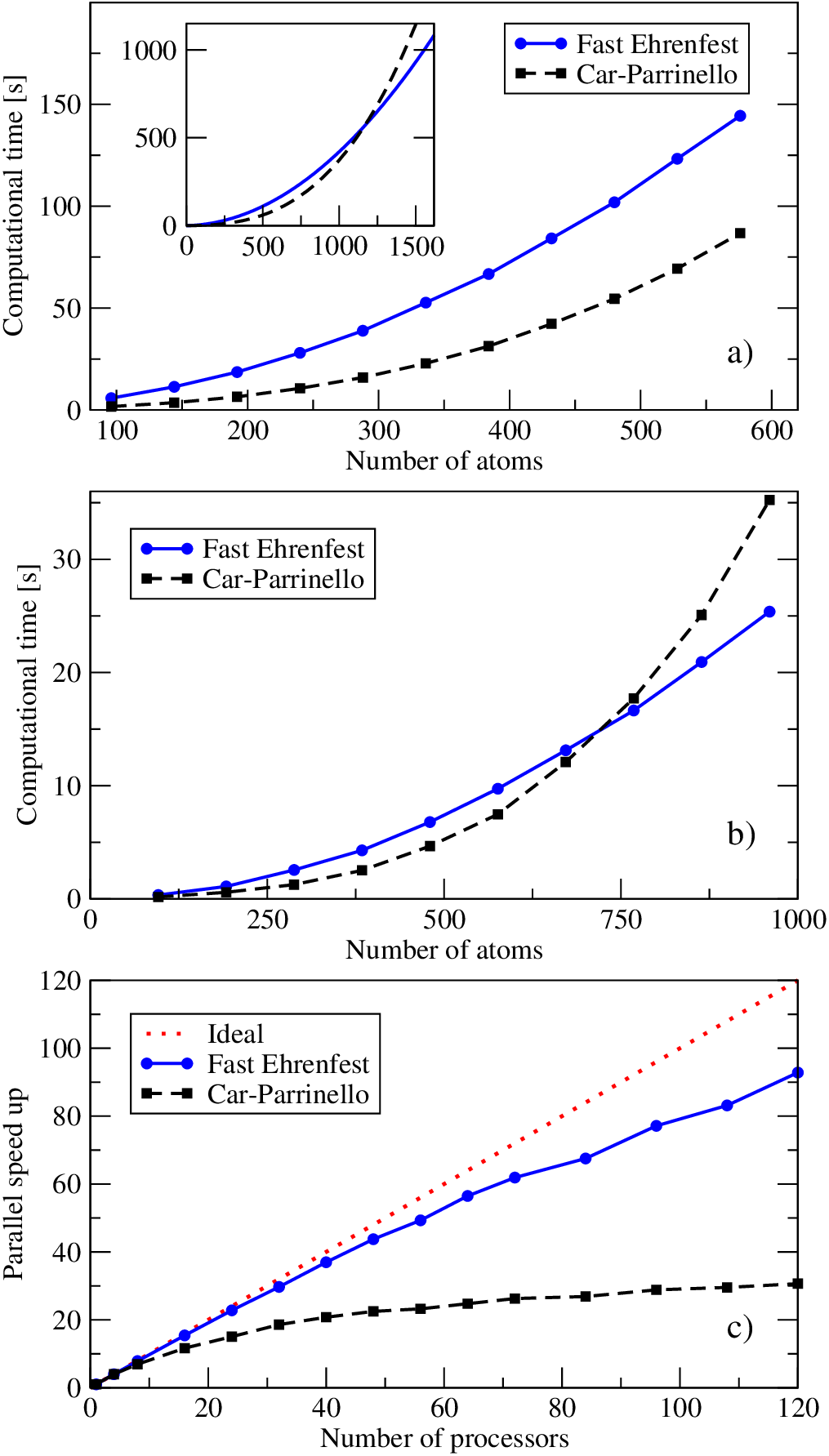}
\caption{\label{performance} 
  Computational performance comparisons of our method and CP for
  an array of benzene molecules with finite boundary conditions
  and a spacing of 0.6 [a.u.]. Performance is measured as the
  computational time required to propagate one atomic unit of time.
  a) Single processor computational cost for different system
  sizes. (inset) Polynomial extrapolation for larger
  systems. Performed in one core of an Intel Xeon E5435 processor.
  b) Parallel computational cost for different system sizes. Performed
  in \(32\,\times\,\)Intel Itanium 2 (1.66 GHz) processor cores of a
  SGI Altix.
  c) Parallel scaling with respect to the number of processor for a
  system of 480 atoms in a SGI Altix system. In both cases a mixed
  states-domain parallelization is used to maximize the performance.
}
\end{figure}

Having established this link, we address the numerical performance of
our new method compared to CP in terms of system size. To do this, we
simulate several benzene molecules in a cell. For the new scheme, a
value of \(\mu=15\) is used while for CP \(\mu_\mathrm{CP}=750\),
(values that yield a similar deviation from the BO surface, according
to Fig.~\ref{benzene}). The time steps used are \(3.15\) [a.u] and
\(7.26\) [a.u.] respectively.  The computational cost is measured as
the simulation time required to propagate one atomic unit of time. We
performed the comparison both for serial and parallel calculations;
the results are shown in Fig.~\ref{performance}. In the serial case,
CP is 3.5 times faster for small systems, but the difference reduces
to only 1.7 times faster for the larger ones. Extrapolating the
results, we predict that the new dynamics will become less demanding
than CP for around 1100 atoms. In the parallel case, the difference is
reduced, CP being only 2 times faster than our method for small
systems, and with a crossing point below 750 atoms. This is due to the
better scalability of the Ehrenfest approach, as seen on
Fig.~\ref{performance}c. Moreover, memory requirements for our
approach are lower than for CP: in the case of 480 atoms the ground
state calculation requires a maximum of 3.5 GB, whereas in the MD,
Ehrenfest requires \(5.6\) GB and CP \(10.5\) GB.

\begin{figure}
\centering
\includegraphics*[scale=0.45]{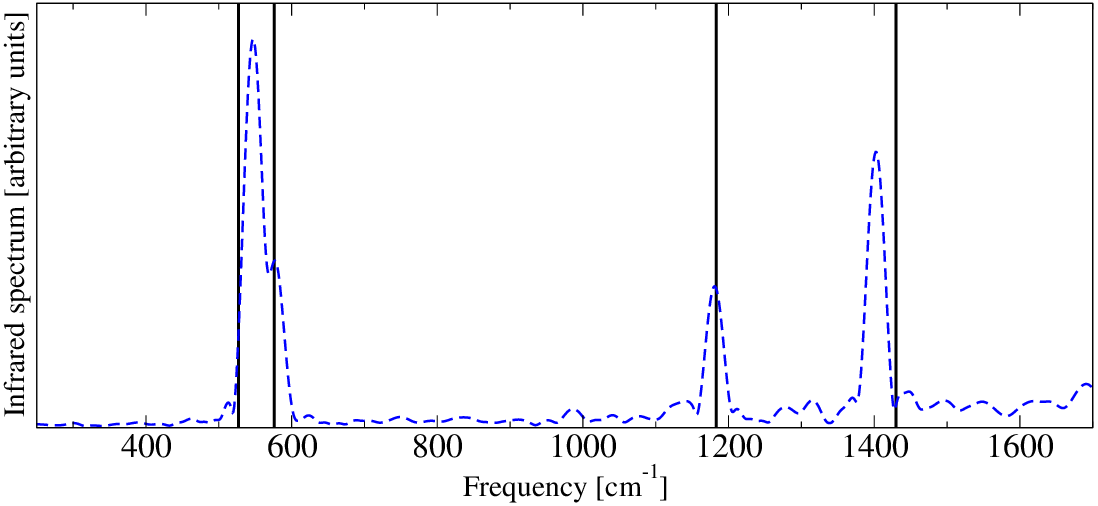}
\caption{\label{fullerenes} Infrared spectrum of C\(_{60}\). The
  (blue) dashed line corresponds to the calculated one (\(\mu\!=\!5\)
  and 2 [ps] of time) while the black bars are the experimental values
  from Ref.~\cite{FullerenesExp}.  }
\end{figure}

To close the computational assessment of the new formalism, we
illustrate our method (using \(\mu=5\)) for the calculation of the
infrared spectrum of a prototype molecule, C\(_{60}\). The calculated
IR spectra is in very good agreement with the experiment (see
Fig.~\ref{fullerenes}) for low and high energy peaks (which are the most
sensitive to \(\mu\) as seen in
Fig.~\ref{benzene}). The result is robust and independent of the
initial condition of the simulation. The low energy splitting of IR
spectrum starts to be resolved for simulations longer than 2 [ps].

In conclusion, we have presented a new approach to AIMD based on a
generalization of TDDFT Ehrenfest dynamics. Our approach introduces a
parameter \(\mu\) that controls the trade-off between the closeness of
the simulation to the gsBO surface and the numerical cost of the
calculation, analogously to the role of the fictitious electronic mass
in CP. We have made direct comparisons of the numerical performance
with CP, and, while quantitatively our results are system- and
implementation-dependent, they prove that our method can outperform CP
in some relevant cases, namely, for large scale systems that are of
interest in several research areas and that can only be studied from
first principles MD in massively parallel computers. To increase its
applicability, it would also be important to study if the improvements
developed to optimize CP can be combined with our
approach~\cite{Kuhne2007}, in particular, techniques to treat small-gap
or metallic systems~\cite{Marzari1997}.

Finally, note that the introduction of the parameter \(\mu\) comes at
a cost, as we change the time scale of the movements of the electrons
with respect to the Ehrenfest case, which implies a shift in the
electronic excitation energies. This must be taken into account when
we extend the applicability of our method for non-adiabatic MD and MD
under electromagnetic fields, in particular for the case of Raman
spectroscopy, general resonant vibrational spectroscopy, and
laser induced molecular bond rearrangement (work in progress).

\begin{acknowledgments}
  We acknowledge support from MEC: FIS2006-12781-C02-01,
  FPA2006-02315, FIS2007-65702-C02-01, BFM2002-00113 and postdoc grant
  for P.~E.; EC: Nanoquanta (NMP4-CT-2004-500198), SANES
  (NMP4-CT-2006-017310) and e-I3 (INFRA-211956); UPV/EHU (SGIker
  Arina) and Barcelona Supercomputing Center. We thank F. Mauri for
  helpful discussions and the referees for suggesting interesting
  ideas to extend the applicability of the method.
\end{acknowledgments}

%\bibliography{aimd}

\end{document}